\documentclass[nohyper,12pt,letterpaper]{JHEP3}
\usepackage{amsfonts,amssymb,amsmath,graphicx}

\newcommand{\be}{\begin{equation}}
\newcommand{\ee}{\end{equation}}
\newcommand{\ben}{\begin{displaymath}}
\newcommand{\een}{\end{displaymath}}
\newcommand{\bea}{\begin{eqnarray}}
\newcommand{\eea}{\end{eqnarray}}
\newcommand{\bean}{\begin{eqnarray*}}
\newcommand{\eean}{\end{eqnarray*}}

\def\l {\lambda}

\def\m{\mu}
\def\cZ{\mathcal{Z}}

\title{\Large Exact solutions for $N$-magnon scattering}

\author{Chrysostomos Kalousios, Georgios Papathanasiou, Anastasia Volovich\\
Department of Physics, Brown University, Box 1843, Providence, RI 02912 \\
E-mail: \email{ckalousi, yorgos, nastja@het.brown.edu}}

\abstract{
Giant magnon solutions play an important role in
various aspects of the AdS/CFT correspondence.
We apply the dressing method to construct an explicit
formula for scattering states of an arbitrary number
$N$ of magnons
on $\mathbb{R} \times S^3$.
The solution can be written in Hirota form and in
terms of determinants of $N \times N$ matrices.
Such a representation may prove useful for the construction
of an effective particle Hamiltonian describing
magnon dynamics.
}


\preprint{\tt{BROWN-HET-1500}}

\begin{document}

\section{Introduction}

Classical string solutions in $AdS_5 \times S^5$ play an important
role in understanding various aspects of the AdS/CFT correspondence
(see \cite{GKP} for review).
Integrability \cite{integrability}
is a powerful computational tool which has
enabled many quantitative checks of the correspondence.
A lot of work has been done exploring both string theory and
gauge theory sides of the correspondence, culminating in the
proposal for an exact S-matrix for planar ${\cal N}=4$
Yang-Mills theory \cite{bes}.

Magnons are building blocks of the spectrum in the spin chain description of AdS/CFT.
The Hofman-Maldacena elementary magnon corresponds to a particular
string configuration moving on an $\mathbb{R} \times S^2$ subspace of
$AdS_5 \times S^5$ \cite{Hofman:2006xt}.
String theory on $\mathbb{R} \times S^2$ (or $\mathbb{R} \times S^3$) is classically
equivalent to sine-Gordon theory (or complex sine-Gordon
theory) via Pohlmeyer reduction \cite{Pohlmeyer:1975nb,mikhailovsinegordon}
(see \cite{devega} for $AdS$ case).
Giant one-magnon solutions on
$\mathbb{R} \times S^2$ and $\mathbb{R} \times S^3$ map to one-soliton solutions in sine-Gordon and
complex sine-Gordon respectively  \cite{Hofman:2006xt,Chen:2006gea}.
Using this map, the scattering phase of two magnons
was computed in \cite{Hofman:2006xt} and shown to match that of \cite{afs}.
Moreover, a sine-Gordon-like
action has been proposed for the
full Green-Schwarz superstring on $AdS_5 \times S^5$
\cite{Grigoriev:2007bu,Mikhailov:2007xr}.

In sine-Gordon theory, the dynamics of $N$-solitons is captured by
the Ruijsenaars-Schneider model \cite{Ruijsenaars:1986vq,Babelon:1993bx}.
Specifically,
the eigenvalues of a particular $N \times N$ matrix entering into the
description of the $N$-soliton solution (or $\tau$-function)
of sine-Gordon evolve according to the Ruijsenaars-Schneider
Hamiltonian. Positions and momenta in the Hamiltonian
are related to the positions and rapidities of the solitons,
and the phase shift for soliton scattering can be calculated from the quantum
mechanical model. It is natural to wonder what the analagous Hamiltonian
in the case of complex sine-Gordon and giant magnons is.
Explicit $N$-soliton solutions (in $\tau$-function form) serve
as a useful starting point in deriving the Ruijsenaars-Schneider
model from the sine-Gordon theory, and it is likely that
a similar technique may prove useful for complex sine-Gordon
and giant magnons as well.

Interest for an effective particle description of giant magnon scattering emerged through the
work of Dorey, Hofman and Maldacena \cite{Dorey:2007xn}, where they illuminated the nature of
double poles appearing in the proposed S-matrix of planar $\mathcal{N}=4$ Yang-Mills \cite{bes}.
They were able to interpret these double poles as occurring from the exchange of
pairs of particles, and in particular to precisely match their position on the complex domain
with the prediction of \cite{bes}, under the assumption that the exchanged particles are
BPS magnon boundstates \cite{Dorey:2006dq}. By studying the quantum mechanical problem
corresponding to an effective particle Hamiltonian describing the scattering of two magnons
with very small relative velocity, one should obtain an S-matrix whose double poles
compare to the aforementioned results in the appropriate limit.

Superposing magnons is a difficult problem because of
the nonlinear equations of motion they satisfy.
Integrability allows the use of algebraic methods such as dressing to construct solutions
of nonlinear equations of motion \cite{Zakharov:1973pp, Zakharov:1980ty}.
Indeed, the dressing method was used to describe the scattering of
two magnons and spikes on
$\mathbb{R} \times S^5$ (and various subsectors)
as well as spikes in  $AdS_3$
\cite{Spradlin:2006wk}--\cite{ Jevicki:2007aa}.
However, it is a tedious process to obtain
even the three-magnon solution.
In this paper we will present an explicit string solution on $\mathbb{R}\times S^3$
describing scattering of an arbitrary number $N$
of magnons by solving the recursive formula
following from the dressing the $(N-1)$-magnon.

The paper is organized as follows. In section 2 we review the dressing method for
$\mathbb{R} \times S^3$ and derive a recursive formula for the $N$-magnon solution in terms of $(N-1)$-magnons.
In section 3 we solve this recursion and present the $N$-magnon solution.
The solution can be presented in various ways, we find useful
Hirota and determinental forms.
As a consistency check we verify that our solution separates
asymptotically into a linear sum of $N$ well-separated single magnon solutions
and demonstrate that the only nontrivial effect of the $N$-magnon interaction
is the expected sum of two-magnon time delays.
The appendix clarifies the rules to construct the $N$-magnon solution and some examples are presented.

\section{Giant magnons on $\mathbb{R} \times S^3$}

The classical action for bosonic strings on $\mathbb{R} \times S^3$ can be written as

\begin{equation}
\label{eq:bosonicaction}
S =
- \frac{1}{ 2 } \int dt\,dx\  \Big[
\partial^a X^\mu \partial_a X_\mu + \Lambda ( X_i \cdot  X_i - 1)\Big],
\end{equation}
where $\mu$ runs from 0 to 4 and $i$ from 1 to 4. The $X_i$ are embedding coordinates on $\mathbb{R}^4$ and the Lagrange multiplier $\Lambda$ constrains them on $S^3$.

After we impose the gauge $X^0(t,x)=t$, eliminate $\Lambda$ in terms of the embedding coordinates and switch to light-cone worldsheet coordinates $z=(x-t)/2, ~\bar{z}=(x+t)/2$, the equations of motion and Virasoro constraints become
\be\label{eom2}
{
\bar{\partial}\partial Z_i+\frac{1}{2}(\partial Z_j\bar\partial \bar{Z}_j+\partial \bar{Z}_j \bar{\partial} Z_j)Z_i=0,~~~Z_i\bar{Z}_i=1
},
\ee
and
\be\label{Virasoro2}
{
\partial Z_i\partial\bar{Z}_i=\bar{\partial} Z_i\bar{\partial}\bar{Z}_i=1
},
\ee
where we have used the parametrization
\begin{equation}
Z_1=X_1+iX_2,~~~Z_2=X_3+iX_4.
\end{equation}

Giant magnons on $\mathbb{R} \times S^3$ are defined as solutions to the above system of equations, obeying the boundary conditions
\begin{eqnarray}\label{bc}
Z_1(t,x\rightarrow\pm \infty)&=&e^{it\pm ip/2+i \alpha}\,,\nonumber\\
Z_2(t,x\rightarrow\pm \infty)&=&0.
\end{eqnarray}
The physical meaning of the boundary conditions is that the endpoints of the string lie on the equator of the $S^3$ on the $Z_1$ plane moving at the speed of light, and the quantity $p$ called total momentum represents the angular distance between them. Finally, $\alpha$ can be any real constant.

\subsection{Review of the dressing method}

The dressing method is a general procedure for constructing soliton solutions to integrable differential equations first developed by Zakharov and Mikhailov \cite{Zakharov:1973pp,Zakharov:1980ty}. It was applied in the context of giant magnons by some of the authors \cite{Spradlin:2006wk,Kalousios:2006xy}, providing
classical solutions for a variety of backgrounds.
In what follows, we will review the basic steps of the method as they apply to the particular case of $\mathbb{R} \times S^3$.

We start by  defining the matrix-valued field
\begin{equation}
\label{SU2}
g(z,\bar{z})\equiv \left( \begin{array}{cc} Z_1 & -iZ_2\\-i\bar{Z}_2 &\bar{Z}_1\\ \end{array}\right)\in SU(2)
\end{equation}
and recasting (\ref{eom2}) into
\begin{equation}
\partial A+\bar{\partial}B=0,
\label{eomdr}
\end{equation}
where the currents $A$ and $B$ are given by
\begin{equation}
\label{currents}
A=i\bar{\partial}g g^{-1}\,,~~~~~B=i\partial g g^{-1}.
\end{equation}
The Virasoro constraints (\ref{Virasoro2}) can be also written as
\begin{equation}
\textrm{Tr} A^2=\textrm{Tr} B^2=2.
\end{equation}

The nonlinear second order equation for $g$ in (\ref{eomdr}) is equivalent to a system of linear first order equations for auxiliary field $\Psi(z,\bar{z},\lambda)$
\begin{equation}
i\partial \Psi=\frac{A \Psi}{1-\lambda}, \qquad i\bar{\partial}\Psi=\frac{B\Psi}{1+\lambda}
\label{drlin}
\end{equation}
provided (\ref{drlin}) holds for any value of the new complex variable $\lambda$ called the spectral parameter, with $A$ and $B$ independent of $\lambda$.

Given any known solution $g$, we can determine $A$, $B$ and solve (\ref{drlin}) to find $\Psi(\lambda)$ subject to the condition
\begin{equation}\label{psicondition}
\Psi(\lambda=0)=g.
\end{equation}
Any ambiguity on factors that don't depend on $z, \bar z$ is removed by also imposing the unitarity condition
\begin{equation}
\left[\Psi(\bar\lambda)\right]^\dag \Psi(\lambda)=I.
\label{unitarity}
\end{equation}
It is easy to show that the equations of motion for the auxiliary field (\ref{drlin}) are covariant under the following transformation with a $\lambda$-dependent parameter $\chi(\lambda)$,
\begin{equation}
\begin{array}{ccccl}
\Psi(\lambda)&\to&\Psi^{\prime}(\lambda)&=&\chi\Psi(\lambda),\\
A&\to&A^{\prime}&=&\chi A\chi^{-1}+i (1+\lambda)\bar\partial\chi\chi^{-1},\\
B&\to&B^{\prime}&=&\chi B\chi^{-1}+i (1-\lambda)\bar\partial\chi\chi^{-1},
\end{array}
\end{equation}
under the condition that $A^\prime, B^\prime$ remain independent of $\lambda$. Thus, performing the above transformation to the known solution $(\Psi(\lambda), A, B)$ produces a new solution to (\ref{eomdr}) with $g^\prime=\Psi^\prime(\lambda=0)$.

The condition (\ref{unitarity}) implies that $\chi(\lambda)$ must obey
\begin{equation}
\left[\chi(\bar\lambda)\right]^\dag\chi(\lambda)=I,
\end{equation}
whereas the demand that $A^\prime, B^\prime$ are independent of $\lambda$ can be translated as further constraints on the analytic properties of $\chi(\lambda)$. For the $\mathbb{R} \times S^3$ case it turns out \cite{Spradlin:2006wk} that the dressing factor $\chi(\lambda)$ is
\begin{equation}
\chi(\lambda)=I+\frac{\lambda_1-\bar{\lambda}_1}{\lambda-\lambda_1} P,
\end{equation}
where $\lambda_1$ is an arbitrary complex number and the hermitian projection operator $P$ is given by
\begin{equation}
P=\frac{\upsilon_1\upsilon_1^{\dagger}}{\upsilon_1^{\dagger} \upsilon_1}, \quad \upsilon_1=\Psi(\bar{\lambda}_1)e,
\end{equation}
where $e$ is an arbitrary vector with constant complex entries called the polarization vector.  The projector $P$ does not depend on the length of the $e$ vector.

The determinant of $\chi(\l)$ is
\be
{\rm det}~\chi(\l)=\frac{\l-\bar{\l}_1}{\l-\l_1}
\ee
and if we want our dressed solution $\chi(0)\Psi(0)$ to sit in $SU(2)$ we should rescale it by the compensating factor $\sqrt{\lambda_1/\bar{\lambda}_1}$.

Putting everything together, the new solution $g^{\prime}=\Psi^{\prime}(\lambda=0)$ to the system (\ref{eomdr}) is given by
\begin{equation}
g^{\prime}=\sqrt{\frac{\lambda_1}{\bar{\lambda}_1}}\left(I+\frac{\lambda_1-\bar{\lambda}_1}{-\lambda_1} P \right)g.
\label{newg}
\end{equation}

\subsection{Application and recursion}

This procedure can be repeated with $g^\prime$ as the solution we begin with, in order to obtain another new solution. In fact, once we have solved the differential equation (\ref{drlin}) for $\Psi(\lambda)$ the first time, we no longer need to repeat this step for $\Psi^\prime(\lambda)$, as we have that information already. Thus, from this point the method proceeds iteratively in a purely algebraic manner.

More specifically, we can show that the auxiliary field $\Psi^N(\l)$ that is constructed after $N$ iterations is related to the auxiliary field $\Psi^{N-1}(\l)$ occuring after $N-1$ iterations through
\be
\label{Psi_N}
\Psi^N(\l)=\sqrt{\frac{\l_N}{\bar{\l}_N}}~\frac{1}{(\l-\l_N)(a b-c d)}\left(
\begin{array}{cc}
\psi_{11}^N & \psi_{12}^N \\
\psi_{21}^N & \psi_{22}^N \\
\end{array}
\right),
\ee
where
\bea
\psi_{11}^N&=&(-c d(\l-\l_N)+a b(\l-\bar{\l}_N))\Psi^{N-1}_{11}(\l)-a c(\l_N-\bar{\l}_N)\Psi^{N-1}_{21}(\l),\nonumber\\
\psi_{12}^N&=&(-c d(\l-\l_N)+a b(\l-\bar{\l}_N))\Psi^{N-1}_{12}(\l)-a c(\l_N-\bar{\l}_N)\Psi^{N-1}_{22}(\l),\nonumber\\
\psi_{21}^N&=&(ab(\l-\l_N)-cd(\l-\bar{\l}_N))\Psi^{N-1}_{21}(\l)+bd(\l_N-\bar{\l}_N)\Psi^{N-1}_{11}(\l),\nonumber\\
\psi_{22}^N&=&(ab(\l-\l_N)-cd(\l-\bar{\l}_N))\Psi^{N-1}_{22}(\l)+bd(\l_N-\bar{\l}_N)\Psi^{N-1}_{12}(\l),
\eea
and
\bea
a&=&\Psi^{N-1}_{11}(\bar{\l}_N)+\Psi^{N-1}_{12}(\bar{\l}_N),\nonumber\\
b&=&\Psi^{N-1}_{21}(\l_N)-\Psi^{N-1}_{22}(\l_N),\nonumber\\
c&=&\Psi^{N-1}_{11}(\l_N)-\Psi^{N-1}_{12}(\l_N),\nonumber\\
d&=&\Psi^{N-1}_{21}(\bar{\l}_N)+\Psi^{N-1}_{22}(\bar{\l}_N).
\eea

The new solution of (\ref{eomdr}) follows from (\ref{Psi_N}) when taking $\lambda=0$. Due to (\ref{SU2}) we can then  read off the relation between the $Z_i$ coordinates of the two solutions as
\begin{eqnarray}
Z^N_1&=&\frac{1}{\vert \l_N\vert (a b-c d)}\left[(a b\bar{\l}_N-c d\l_N)Z_1^{N-1}+a c(\l_N-\bar{\l}_N)(-i \bar Z_2^{N-1})\right],\nonumber\\
Z^N_2&=&\frac{i}{\vert \l_N\vert (a b-c d)}\left[(a b\bar{\l}_N-c d\l_N)(-i Z_2^{N-1})+a c(\l_N-\bar{\l}_N)\bar Z_1^{N-1}\right].
\end{eqnarray}

Starting with the simple `vacuum' solution representing a point particle rotating around the equator in the $Z_1$ plane,
\begin{eqnarray}
Z_1&=&e^{it}\,,\nonumber\\
Z_2&=&0,
\end{eqnarray}
and using the polarization vector $e=(1,~1)$ the dressing method yields \cite{Spradlin:2006wk} the single magnon solution on $\mathbb{R}\times S^3$ first obtained in \cite{Chen:2006gea} as a generalization of the original Hofman-Maldacena giant magnon solution on $\mathbb{R}\times S^2$. Applying the method once more using the same polarization vector as before then gives a solution which asymptotically reduces to a sum of two single magnon solutions, and whose conserved charges are sums of the respective charges of two single magnon solutions. Hence it can be interpreted as a scattering state of two single magnons.

{}From the above considerations, it is natural to expect that the $N$-times dressed solution will correspond to a scattering state of $N$ magnons. The quantities $\lambda_i$ are parameters of the $N$-magnon solution which we can more conventionally  express as $\lambda_i=r_i e^{ip_i/2}$, with $p_i$ the momentum of each constituent magnon and $r_i$ a quantity associated to its $U(1)$ charge.

\section{The $N$-magnon solution}
Successive application of the dressing method suggests a compact closed form for the $N$-magnon solution, which can be written as follows
\begin{eqnarray}\label{Z1}
Z_1&=&\frac{e^{it}}{\prod_{l=1}^{N}|\lambda_l|}\frac{N_{1}}{D}\,,\nonumber\\
Z_2&=&-i\frac{e^{-it}}{\prod_{l=1}^{N}|\lambda_l|}\frac{N_{2}}{D},
\end{eqnarray}
with
\bea
\label{nmagfirst}
D&=&\sum_{\m_i=0,1}\exp\left[\sum_{i<j}^{2N}B_{ij}[\m_i\m_j+(\m_i-1)(\m_j-1)]+\sum_{i=1}^{2N}\m_i (2i \mathcal{Z}_i)\right],\nonumber\\
N_1&=&\sum_{\m_i=0,1}\exp\left[\sum_{i<j}^{2N}B_{ij}[\m_i\m_j+(\m_i-1)(\m_j-1)]+\sum_{i=1}^{2N}\m_i (2i \mathcal{Z}_i+C_i)\right],\\
N_2&=&\sum_{\m_i=0,1}\exp\left[\sum_{i<j}^{2N}B_{ij}[\m_i\m_j+(\m_i-1)(\m_j-1)]+\sum_{i=1}^{2N}[\m_i (2i \mathcal{Z}_i)+(\m_i-1)C_i]\right]\nonumber,
\eea
where
\begin{eqnarray}
\mathcal{Z}_i&=&\frac{z}{\l_i-1}+\frac{\bar{z}}{\l_i+1},\nonumber\\
e^{B_{ij}}&=&\l_{i}-\l_j,\\
e^{C_i}&=&\l_i,\nonumber
\end{eqnarray}
and $N$ is the number of magnons.

In the above formula the indices $i,j$ take the $2N$ values $(1,\bar 1,2,\bar 2, ...,\bar N)$, $i<j$ implies this particular ordering, and we identify $\lambda_{\bar k}\equiv\bar \lambda_k, ~ \mathcal{Z}_{\bar k}\equiv \mathcal{\bar Z}_k$ \footnote{Alternatively we may define new quantities $\rho_k$ such that $\rho_{2l-1}=\l_l$ and $\rho_{2l}=\bar\l_l$, and similarly for $\mathcal{Z}_l$. These will take values $1,2...2N$ as usual.}. The symbol $\sum_{\m_i=0,1}$ implies the summation over all possible combinations of $\mu_1=0, 1,~\mu_{\bar{1}}=0,1,\ldots,~\mu_{\bar N}=0,1$ under the conditions
\be
\sum_{i=1}^{2N} \m_i=
\begin{cases}
N, & {\rm for ~} N_1,~D,\\
N+1, & {\rm for ~} N_2.
\label{nmaglast}
\end{cases}
\ee
This description makes contact with a variety of $N$-soliton expressions of other integrable systems (for example see \cite{Scott:1973eg}).

We have numerically checked (\ref{nmagfirst}) for high number of magnons, whereas in fig. \ref{plot} we plot $|Z_2|$ for the first 4 magnons.  In appendix \ref{appA} we give some examples.

Our $\mathbb{R} \times S^3$ $N$-magnon solution is reduced to the $\mathbb{R} \times S^2$ one if we let the spectral parameters $\l_l$ lie on a unit circle, $|\l_l|=1$.

\begin{figure}[h!]
\begin{center}
\includegraphics[width=0.9\textwidth]{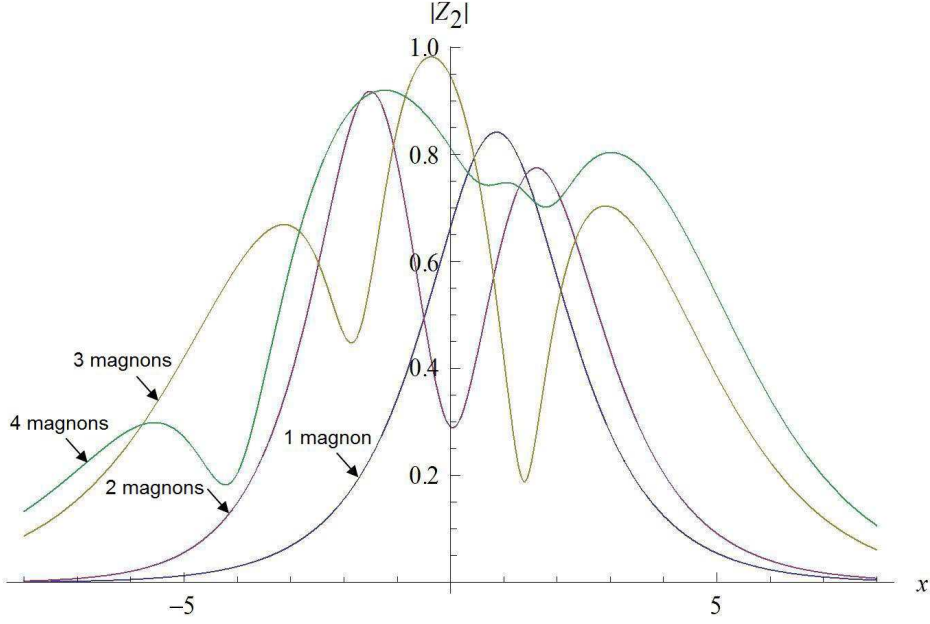}
\caption{Plot of $|Z_2|$ for the first 4 magnons on $\mathbb{R} \times S^3$ at time t=2 as a function of the worldsheet coordinate $x$.  The chosen spectral parameters are $\l_1=2e^{i},~\l_2=e^{2i},~\l_3=3e^{2i},~\l_4=e^{4i}$.}
\label{plot}
\end{center}
\end{figure}

\subsection{Hirota form of the solution}
It is possible to write $Z_1,~Z_2$ of (\ref{Z1}) in an equivalent form similar to Hirota's \cite{1972JPSJ...33.1459H}, where $N_1,~N_2,~D$ are given by
\bea\label{Hirota1}
D&=&\sum_{_{2N}C_N} d(i_1,i_2,\ldots,i_N)  \exp\left[ 2 i (\mathcal{Z}_{i_1}+\mathcal{Z}_{i_2}+\cdots+\mathcal{Z}_{i_N})\right],\nonumber\\
N_1&=&\sum_{_{2N}C_N} n_1(i_1,i_2,\ldots,i_N)  \exp\left[2 i (\mathcal{Z}_{i_1}+\mathcal{Z}_{i_2}+\cdots+\mathcal{Z}_{i_N})\right], \\\nonumber
N_2&=&\sum_{_{2N}C_{N+1}} n_2(i_1,i_2,\ldots,i_{N+1}) \exp\left[2 i  (\mathcal{Z}_{i_1}+\mathcal{Z}_{i_2}+\cdots+\mathcal{Z}_{i_{N+1}})\right],
\eea
and
\bea\label{Hirota2}
d(i_1,i_2,\ldots,i_N)&=&\prod_{k<l\leq N}^{(N)}\l_{i_k i_l} \prod_{N<m<n}^{(N)}\l_{i_m i_n},\nonumber\\
n_1(i_1,i_2,\ldots,i_N)&=&\prod_{j=1}^{N} \l_{i_j} \prod_{k<l\leq N}^{(N)}\l_{i_k i_l} \prod_{N<m<n}^{(N)}\l_{i_m i_n},\\\nonumber
n_2(i_1,i_2,\ldots,i_{N+1})&=&\prod_{j=N+1}^{2N} \l_{i_j} \prod_{k<l\leq N+1}^{(N+1)}\l_{i_k i_l} \prod_{N+1<m<n}^{(N-1)}\l_{i_m i_n},
\label{nmagnonlast}
\eea
where $N$ is the number of magnons, $_{N}C_n$ indicates summation over all possible combinations of $n$ elements taken from $N$, $\prod^{(n)}$ indicates the product of all possible combinations of the $n$ elements, and $\l_{ij}=\l_i-\l_j$.  Finally, we have arranged our $2N$ elements $\mathcal{Z}_i$ as $\{\mathcal{Z}_1,\bar{\mathcal{Z}}_1,\ldots,\bar{\mathcal{Z}
}_N\}$ and our $2N$ $\l$'s as $\{\mathcal{\l}_1,\bar{\mathcal{\l}}_1,\ldots,\bar{\mathcal{\l}}_N\}$.  We always assume that $i_1<\ldots <i_N$.

Finally, we should mention that we can get a more symmetric yet complicated-looking version of our $N$-magnon expressions, by factoring out the terms


\be\label{factorout}
\begin{cases}
\prod_{l=1}^N \l_l\exp \left(2i\sum_{l=1}^N \mathcal{Z}_l \right) &\textrm{from $N_1$,}\\
\prod_{l=1}^N \bar\l_l\exp \left(2i\sum_{l=1}^N \mathcal{Z}_l \right) &\textrm{from $N_2$,}\\
\exp\left(2i\sum_{l=1}^N \mathcal{Z}_l \right) &\textrm{from $D$.}
\end{cases}
\ee

Written in this way, $D$ has the nice feature of being real. More importantly, and as we will see in the following sections, this form of the $N$-magnon solution is useful for analyzing its asymptotic behavior and demonstrates the symmetry that will allow us to write it in a determinant form.

\subsection{Determinant form for $Z_1$}

It is known that for the (complex) sine-Gordon equation and several other integrable equations, the $N$-soliton expressions similar to (\ref{Z1})-(\ref{nmaglast}) and (\ref{Hirota1})-(\ref{Hirota2}) can also be rewritten in a form involving determinants of $N\times N$ matrices \cite{deVega:1982sh}. It is precisely expressions of this type that become particularly useful when extracting the effective particle description of the soliton problem \cite{Babelon:1993bx}. Motivated by the same goal for the case of giant magnons, we haven been able to find a determinant formula for $Z_1$. In particular, we may write
\be
\label{det}
Z_1=e^{it}\prod^N_{l=1}\left(\frac{\l_l}{\bar{\l}_l}\right)^{1/2}\frac{{\rm det}(I+\Lambda^{-1}F\bar\Lambda \bar F)}{{\rm det}(I+F\bar F)},
\ee
where $\Lambda, F$ are $N\times N$ matrices\footnote{The matrix $\Lambda$ is not to be confused with the Lagrange multiplier of (\ref{eq:bosonicaction}).} with elements
\begin{eqnarray}
\Lambda_{kl}&=&\delta_{kl} \lambda_l\nonumber,\\
F_{kl}&=&e^{-2i\mathcal{Z}_k}G_{kl},\\
G_{kl}&=&\prod_{m\neq l}\frac{\l_{k\bar{m}}}{\l_{\bar{l}\bar{m}}}\nonumber,
\end{eqnarray}
$k,l=1,2,\ldots,N$, and $I$ the identity matrix. Interestingly, the matrix $G$ can further be expressed as $G=H \left(\bar H\right)^{-1}$ where $H$ is a matrix with elements $H_{kl}=(\lambda_k)^{l-1}$. The determinant of $H$ is what is known in the literature as the Vandermonde determinant, given by the simple formula
\begin{equation}
\det H=\prod_{k<l} (\l_l-\l_k).
\end{equation}
This decomposition in terms of $H$ also reveals the property of $G$, that $\bar G=G^{-1}$. Finally, one may use the property that two square matrices related by a similarity transformation $A^\prime=S A S^{-1}$ obey $\det(I+A^\prime)=\det(I+A)$ to regroup the matrix products of (\ref{det}) in a different manner if desired.

The fact that the exponents in $N_2$ contain $N+1$ $\mathcal{Z}_i$ terms complicates the derivation of a determinant formula for $Z_2$.

\subsection{Asymptotic behavior}

In this section we will examine how our solution behaves for $x\to\pm\infty$ and $t\to\pm\infty$ respectively. Since the dependence of our solutions on
the worldsheet coordinates is encoded in the factors $2 i \mathcal{Z}_i$, the asymptotic behavior of the $N$-magnon solution will be determined by their respective real parts.

Using notation similar to \cite{Spradlin:2006wk}, we define
\bea
u_l&\equiv& i (\mathcal{Z}_l-\bar{\mathcal{Z}}_l)=
\kappa_l x- \nu_l t,\nonumber\\
w_l&\equiv& \mathcal{Z}_l+\bar{\mathcal{Z}}_l,\nonumber\\
v_l&\equiv& w_l-t,
\eea
with
\bea
\label{uw}
\kappa_l&=&-i\frac{ (\l_l-\bar\l_l)(1+|\l_l|^2)}{\left|1-\l_l\right|^2\left|1+\l_l\right|^2}=\frac{2(1+r_l^2)r_l\sin \frac{p_l}{2}}{1+r_l^4-2r_l^2\cos p_l},\nonumber\\
\nu_l&=&\frac{-i(\l_l^2-\bar\l_l^2)}{\left|1-\l_l\right|^2\left|1+\l_l\right|^2}=\frac{2r_l \sin p_l}{1+r_l^4-2r_l^2\cos p_l},
\eea
and in the second equality we have also employed the usual parametrization $\l_l=r_l e^{ip_l/2}$ for the spectral parameters. Additionally, the relations (\ref{uw}) imply
\be
2 i  \mathcal{Z}_l=u_l+ i w_l,~~~2 i \bar{\mathcal{Z}}_l=-u_l+ i w_l.
\ee

The parameter range for a single dyonic magnon is $r\in(0,\infty)$ and $p\in\left[0,2\pi \right)$, with $p\sim p+2\pi$ for any other $p$. We can use the same restrictions for our parameters $r_l, p_l$ of the $N$-magnon solution, in which case the $\kappa_l$ are clearly positive. {}From the formulas
(\ref{Hirota1})-(\ref{Hirota2}) after we factor out (\ref{factorout}), it is then easy to see that the our solution has its boundaries on the equator of $S^3$ on the $Z_1$ plane. Namely, for $x\to\pm\infty$ the boundary conditions (\ref{bc}) are satisfied, with $p=\sum_{l=1}^{N}p_l$ as expected.

Next, we proceed to determine the behavior of the solution for $t\to\pm\infty$ and large magnon separation. Without loss of generality, we can assume that the magnons are ordered such that their velocities $\frac{\nu_k}{\kappa_k}$ obey
\be
\frac{\nu_1}{\kappa_1}>\frac{\nu_2}{\kappa_2}>...>\frac{\nu_N}{\kappa_N}.
\ee

In order to focus on the $k$-th magnon, we keep $u_k$ fixed as $t\to\pm\infty$. This means that $x$ should scale as $x=\frac{\nu_k}{\kappa_k} t+\frac{u_k}{\kappa_k}$ and in total the $u_l$ will behave as
\be
u_l=\kappa_l \left(\frac{\nu_k}{\kappa_k}-\frac{\nu_l}{\kappa_l}\right)t+\kappa_l\frac{u_k}{\kappa_k}.
\ee
In particular, the limit $t\to-\infty$ under the aforementioned ordering and scaling implies
\bea
&&u_1, u_2,\ldots,u_{k-1}\to+\infty\nonumber,\\
&&u_k ~~\text{finite},\\
&&u_{k+1},u_{k+2},\ldots,u_N\to-\infty\nonumber.
\eea
Thus, it is easy to see from (\ref{nmagfirst})-(\ref{nmaglast}) that the terms which dominate in the limit have $\mu_i=1$ for $i\in\{1,\ldots,k-1,k,\overline{k+1},\ldots,\overline{N}\}$ and $i\in\{1,\ldots,k-1,\overline{k},\overline{k+1},\ldots,\overline{N}\}$ in the case of $N_1, D$, and $i\in\{1,\ldots,k-1,k,\overline{k},\overline{k+1},\ldots,\overline{N}\}$ in the case of $N_2$, with the rest of the $\mu$'s being zero.

Up to common factors that will eventually cancel out (including the divergent terms), we can express the limiting values of $N_1, N_2~\text{and}~D$ as
\bea
\label{limit1}
D&\sim&\left(f_{+}e^{u_k}+f_{-}e^{-u_k}\right)e^{i w_k},\nonumber\\
N_1&\sim&\prod_{l=1}^{k-1}\l_l\prod_{l=k+1}^{N}\bar\l_l~\left(\l_k \,f_{+}\,e^{u_k}+\bar \l_k\, f_{-}\,e^{-u_k}\right)e^{i w_k},\\
N_2&\sim&\prod_{l=1}^{k-1}\bar\l_l\prod_{l=k+1}^{N}\l_l~ \l_{2\bar2}~h ~e^{2 i w_k},\nonumber
\eea
where $f_+,~f_-,~h$ are functions of the spectral parameters $\l_i$ given by
\bea
\label{limit2}
f_+&=&\prod_{l=1}^{k-1}\vert\l_k-\l_l\vert^2\prod_{l=k+1}^{N}\vert\bar\l_k-\l_l\vert^2,\nonumber\\
f_-&=&\prod_{l=1}^{k-1}\vert\bar \l_k-\l_l\vert^2\prod_{l=k+1}^{N}\vert\l_k-\l_l\vert^2,\\
h&=&\prod_{l=1}^{k-1}(\l_k-\l_l)(\bar\l_k-\l_l)\prod_{l=k+1}^{N}(\l_k-\bar\l_l)(\bar \l_k-\bar\l_l)\nonumber.
\eea
Noticing that $\vert h\vert^2=f_+f_-$, and with the help of (\ref{Z1}), (\ref{limit1}) and (\ref{limit2}), we can write the $t\to-\infty$ limit of the $N$-magnon solution as
\bea
\label{limit3}
Z_1&=&e^{i\theta_1} e^{i t}\left[\cos\frac{p_k}{2}+i \sin\frac{p_k}{2}\tanh (u_k+\delta u_-(k))\right],\nonumber\\
Z_2&=&e^{i\theta_2}e^{i v_k}\frac{\sin \frac{p_k}{2}}{\cosh\left[u_k+\delta u_-(k)\right]},
\eea
where\footnote{The signs of $\delta u_\pm(k)$ are chosen for compatibility with the most standard method of determining time delays, whereby one performs the ansatz $u_k=-\nu_k \delta t_\pm(k)$ and solves for the position of the magnon's peak, given by $-\nu_k \delta t_\pm(k)+\delta u_\pm(k)=0$. Note the agreement with the definition (\ref{deltatotal}) below.}
\be
\label{deltaplus}
\delta u_-(k)=\frac{1}{2}\log \frac{f_+}{f_-}=\sum_{l=1}^{k-1}\delta u_{k,l}-\sum_{l=k+1}^{N}\delta u_{k,l}
\ee
with
\be
\label{deltalk}
\delta u_{k,l}=\log \left| \frac{\l_k-\l_l}{\bar\l_k-\l_l}\right|,
\ee
and the phase factors $e^{i\theta_1}, ~ e^{i\theta_2}$ are independent of $x$ and $t$. For completeness, we can write them explicitly as
\bea
e^{i \theta_1}&=&\prod_{l=1}^{k-1}\left(\frac{\l_l}{\bar{\l_l}}\right)^{1/2}\prod_{l=k+1}^{N}\left(\frac{\bar\l_l}{\l_l}\right)^{1/2} =\exp\left[\frac{i}{2} \left( \sum_{l=1}^{k-1}p_l-\sum_{l=k+1}^{N}p_l\right)\right],\nonumber\\
e^{i \theta_2}&=&e^{i\zeta}e^{-i\theta_1}=\left(\frac{h}{\bar h}\right)^{1/2}\,e^{-i\theta_1}.
\eea
Equation (\ref{limit3}) is precisely the single magnon solution on $\mathbb{R}\times S^3$ \cite{Chen:2006gea, Spradlin:2006wk}, up to a pure phase and a shift in $u_k$, which reflects the additional freedom of the solution.

The case $t\to\infty$ can be treated in a similar manner, yielding (\ref{limit3}) with
\bea
\delta u_-(k)&\to&\delta u_+(k)=-\delta u_-(k),\nonumber\\
\theta_1&\to&-\theta_1,\\
\zeta&\to&-\zeta.\nonumber\\
\eea
Since $k$ is arbitrary, we have in fact proven that asymptotically our $N$-magnon solution splits into $N$ single magnon solutions. Each magnon retains its shape after scattering with the rest of the magnons, with the effect of the interaction being encoded only in a relative shift in $u_k$,
\be
\delta u(k)\equiv\delta u_+(k)-\delta u_-(k)=-2\delta u_-(k).
\ee
Because of (\ref{uw}), the shift in $u_k$ is usually interpreted as a time delay \cite{Jackiw:1975im},
\be
\label{deltatotal}
\delta t(k)\equiv\frac{\delta u(k)}{\nu_k}=-\sum_{l=1}^{k-1}\delta t_{k,l}+\sum_{l=k+1}^{N}\delta t_{k,l},
\ee
where
\be
\label{timedelay}
\delta t_{k,l}\equiv\frac{2\delta u_{k,l} }{\nu_k}=2i\frac{\left|1-\l_k\right|^2\left|1+\l_k\right|^2}{\l_k^2-\bar\l_k^2}\log \left| \frac{\l_k-\l_l}{\bar\l_k-\l_l}\right|
\ee
is the time delay that occurs because of the interaction of the $k$-th with the $l$-th magnon, namely two-magnon scattering.

Hence, our $N$-magnon solution exhibits the property of factorized scattering, as expected by the integrability of the $\mathbb{R}\times S^3$ $\sigma$-model and its classical equivalence to the complex sine-Gordon system. Finally, the dyonic two-magnon time-delay we retrieved in (\ref{timedelay}) is in complete agreement with \cite{Chen:2006gq,Roiban:2006gs,Kalousios:2006xy}.

\section*{Acknowledgments}

We are grateful to M. Abbott, I. Aniceto, J. Avan and K. Jin,
and especially to A. Jevicki and M. Spradlin, for comments and discussions. This work is supported by DOE grant DE-FG02-91ER40688. The research of AV is also supported by NSF CAREER Award PHY-0643150.

\appendix
\section{Construction rules - Examples}\label{appA}
To help clarify the meaning of the formulas (\ref{Z1})-(\ref{nmaglast}) and (\ref{Hirota1})-(\ref{Hirota2}), we reduce them to a simple set of rules for the construction of $N_{1},~N_{2},~D$. These rules may also facilitate computer code for generating $N$-magnon solutions.

The $N$-magnon solution can be written as
\begin{eqnarray}
Z_1&=&\frac{e^{it}}{\prod_{l=1}^{N}|\lambda_l|}\frac{N_{1}}{D}\,,\nonumber\\
Z_2&=&-i\frac{e^{-it}}{\prod_{l=1}^{N}|\lambda_l|}\frac{N_{2}}{D}.
\end{eqnarray}
and it contains $N$ spectral parameters $\lambda_i$ along with their conjugates $\bar{\lambda}_i$ that we can arrange as the set $A=\{\lambda_1,~\bar{\lambda}_1,~\lambda_2,\ldots,~\lambda_N,~\bar{\lambda}_N\}$.

In order to write the denominator $D$ we take all the possible subsets of $N$ numbers of the set $A$. There are $(2N)!/N!^2$ such subsets.  For each subset we form a product and then D is the sum of all those products.  Let us see how to form the product for a specific subset $B$.  The product contains

a) an exponential with exponent $2i \sum_i \cZ(\l_i) \equiv 2i \sum_i \cZ_i$, where $\lambda_i$ are all the $\lambda$'s that belong to $B$,

b) all the possible differences $\lambda_i-\lambda_j,~i<j$, where $\lambda_i, ~\lambda_j$ all belong to the subset B and

c) finally all the possible differences $\lambda_i-\lambda_j,~i<j$, where $\lambda_i, ~\lambda_j$ all belong to the complement subset of $B$.

The rules for $N_1$ are the same as $D$ except that now the product contains in addition all the $\lambda$'s that belong to the subset $B$.

The rules for $N_2$ are the same as the rules for $N_1$, but now all the subsets $B$ should have $N+1$ elements instead of $N$ and the product contains all the $\lambda$'s that belong to the complement subset of $B$ instead of the $B$ itself.

As an example let us write $N_{1},~N_{2},~D$ in the case of 1, 2 and 3-magnons.\\
For 1-magnon we have \cite{Chen:2006gea}
\bea
D&=&e^{2i\cZ_1}+e^{2i\bar{\cZ}_1},\nonumber\\
N_1&=&\l_1 e^{2i\cZ_1}+\bar{\l}_1e^{2i\bar{\cZ}_1},\\\nonumber
N_2&=&\l_{1\bar{1}}e^{2i(\cZ_1+\bar{\cZ}_1)}.
\eea
For 2-magnons we have \cite{Spradlin:2006wk}
\begin{eqnarray}
D&=&\lambda_{1\bar{1}}\lambda_{2\bar{2}}e^{2i(\cZ_1+\bar{\cZ}_1)} +\lambda_{12}\lambda_{\bar{1}\bar{2}}e^{2i(\cZ_1+\cZ_2)} +\lambda_{1\bar{2}}\lambda_{\bar{1}2}e^{2i(\cZ_1+\bar{\cZ}_2)} \nonumber\\
&&  +\lambda_{\bar{1}2}\lambda_{1\bar{2}}e^{2i(\bar{\cZ}_1+\cZ_2)} +\lambda_{\bar{1}\bar{2}}\lambda_{12}e^{2i(\bar{\cZ}_1+\bar{\cZ}_2)} +\lambda_{2\bar{2}}\lambda_{1\bar{1}}e^{2i(\cZ_2+\bar{\cZ}_2)},\nonumber\\
N_1&=&\lambda_1\bar{\lambda}_1\lambda_{1\bar{1}}\lambda_{2\bar{2}}e^{2i(\cZ_1+\bar{\cZ}_1)} +\lambda_1\lambda_2 \lambda_{12}\lambda_{\bar{1}\bar{2}}e^{2i(\cZ_1+\cZ_2)} +\lambda_1\bar{\lambda}_2\lambda_{1\bar{2}}\lambda_{\bar{1}2}e^{2i(\cZ_1+\bar{\cZ}_2)} \nonumber\\
&&  ~+\bar{\lambda}_1\lambda_2\lambda_{\bar{1}2}\lambda_{1\bar{2}}e^{2i(\bar{\cZ}_1+\cZ_2)} +\bar{\lambda}_1\bar{\lambda}_2\lambda_{\bar{1}\bar{2}}\lambda_{12}e^{2i(\bar{\cZ}_1+\bar{\cZ}_2)} +\lambda_2\bar{\lambda}_2\lambda_{2\bar{2}}\lambda_{1\bar{1}}e^{2i(\cZ_2+\bar{\cZ}_2)},\nonumber\\
N_2&=&\bar{\l}_2\l_{1\bar{1}}\l_{12}\l_{\bar{1}2}e^{2i(\cZ_1+\bar{\cZ}_1+\cZ_2)} +\l_2\l_{1\bar{1}}\l_{1\bar{2}}\l_{\bar{1}\bar{2}}e^{2i(\cZ_1+\bar{\cZ}_1+\bar{\cZ}_2)}\nonumber\\
&&  +\bar{\l}_1\l_{12}\l_{1\bar{2}}\l_{2\bar{2}}e^{2i(\cZ_1+\cZ_2+\bar{\cZ}_2)} +\l_1\l_{\bar{1}2}\l_{\bar{1}\bar{2}}\l_{2\bar{2}}e^{2i(\bar{\cZ}_1+\cZ_2+\bar{\cZ}_2)}.
\end{eqnarray}
For 3-magnons we have
\begin{equation*}
\begin{split}
D&=\l_{1\bar{1}}\l_{12}\l_{\bar{1}2}\l_{\bar{2}3}\l_{\bar{2}\bar{3}}\l_{3\bar{3}}e^{2i(\cZ_1+\bar{\cZ}_1+\cZ_2)} +\l_{1\bar{1}}\l_{1\bar{2}}\l_{\bar{1}\bar{2}}\l_{23}\l_{2\bar{3}}\l_{3\bar{3}}e^{2i(\cZ_1+\bar{\cZ}_1+\bar{\cZ}_2)} \\
& \quad +\l_{1\bar{1}}\l_{13}\l_{\bar{1}3}\l_{2\bar{2}}\l_{2\bar{3}}\l_{\bar{2}\bar{3}}e^{2i(\cZ_1+\bar{\cZ}_1+\cZ_3)} +\l_{1\bar{1}}\l_{1\bar{3}}\l_{\bar{1}\bar{3}}\l_{2\bar{2}}\l_{23}\l_{\bar{2}3}e^{2i(\cZ_1+\bar{\cZ}_1+\bar{\cZ}_3)} \\
& \quad +\l_{12}\l_{1\bar{2}}\l_{2\bar{2}}\l_{\bar{1}3}\l_{\bar{1}\bar{3}}\l_{3\bar{3}}e^{2i(\cZ_1+\cZ_2+\bar{\cZ}_2)} +\l_{12}\l_{13}\l_{23}\l_{\bar{1}\bar{2}}\l_{\bar{1}\bar{3}}\l_{\bar{2}\bar{3}}e^{2i(\cZ_1+\cZ_2+\cZ_3)} \\
& \quad +\l_{12}\l_{1\bar{3}}\l_{2\bar{3}}\l_{\bar{1}\bar{2}}\l_{\bar{1}3}\l_{\bar{2}3}e^{2i(\cZ_1+\cZ_2+\bar{\cZ}_3)} +\l_{1\bar{2}}\l_{13}\l_{\bar{2}3}\l_{\bar{1}2}\l_{\bar{1}\bar{3}}\l_{2\bar{3}}e^{2i(\cZ_1+\bar{\cZ}_2+\cZ_3)} \\
& \quad +\l_{1\bar{2}}\l_{1\bar{3}}\l_{\bar{2}\bar{3}}\l_{\bar{1}2}\l_{\bar{1}3}\l_{23}e^{2i(\cZ_1+\bar{\cZ}_2+\bar{\cZ}_3)}
+\l_{13}\l_{1\bar{3}}\l_{3\bar{3}}\l_{\bar{1}2}\l_{\bar{1}\bar{2}}\l_{2\bar{2}}e^{2i(\cZ_1+\cZ_3+\bar{\cZ}_3)} \\
& \quad +\l_{\bar{1}2}\l_{\bar{1}\bar{2}}\l_{2\bar{2}}\l_{13}\l_{1\bar{3}}\l_{3\bar{3}}e^{2i(\bar{\cZ}_1+\cZ_2+\bar{\cZ}_2)} +\l_{\bar{1}2}\l_{\bar{1}3}\l_{23}\l_{1\bar{2}}\l_{1\bar{3}}\l_{\bar{2}\bar{3}}e^{2i(\bar{\cZ}_1+\cZ_2+\cZ_3)} \\
& \quad +\l_{\bar{1}2}\l_{\bar{1}\bar{3}}\l_{2\bar{3}}\l_{1\bar{2}}\l_{13}\l_{\bar{2}3}e^{2i(\bar{\cZ}_1+\cZ_2+\bar{\cZ}_3)} +\l_{\bar{1}\bar{2}}\l_{\bar{1}3}\l_{\bar{2}3}\l_{12}\l_{1\bar{3}}\l_{2\bar{3}}e^{2i(\bar{\cZ}_1+\bar{\cZ}_2+\cZ_3)} \\
& \quad +\l_{\bar{1}\bar{2}}\l_{\bar{1}\bar{3}}\l_{\bar{2}\bar{3}}\l_{12}\l_{13}\l_{23}e^{2i(\bar{\cZ}_1+\bar{\cZ}_2+\bar{\cZ}_3)} +\l_{\bar{1}3}\l_{\bar{1}\bar{3}}\l_{3\bar{3}}\l_{12}\l_{1\bar{2}}\l_{2\bar{2}}e^{2i(\bar{\cZ}_1+\cZ_3+\bar{\cZ}_3)} \\
& \quad +\l_{2\bar{2}}\l_{23}\l_{\bar{2}3}\l_{1\bar{1}}\l_{1\bar{3}}\l_{\bar{1}\bar{3}}e^{2i(\cZ_2+\bar{\cZ}_2+\cZ_3)} +\l_{2\bar{2}}\l_{2\bar{3}}\l_{\bar{2}\bar{3}}\l_{1\bar{1}}\l_{13}\l_{\bar{1}3}e^{2i(\cZ_2+\bar{\cZ}_2+\bar{\cZ}_3)} \\
& \quad +\l_{23}\l_{2\bar{3}}\l_{3\bar{3}}\l_{1\bar{1}}\l_{1\bar{2}}\l_{\bar{1}\bar{2}}e^{2i(\cZ_2+\cZ_3+\bar{\cZ}_3)} +\l_{\bar{2}3}\l_{\bar{2}\bar{3}}\l_{3\bar{3}}\l_{1\bar{1}}\l_{12}\l_{\bar{1}2}e^{2i(\bar{\cZ}_2+\cZ_3+\bar{\cZ}_3)},
\end{split}
\end{equation*}

\begin{equation*}
\begin{split}
N_1&=\l_1\bar{\l}_1\l_2\l_{1\bar{1}}\l_{12}\l_{\bar{1}2}\l_{\bar{2}3}\l_{\bar{2}\bar{3}}\l_{3\bar{3}}e^{2i(\cZ_1+\bar{\cZ}_1+\cZ_2)} +\l_1\bar{\l}_1\bar{\l}_2\l_{1\bar{1}}\l_{1\bar{2}}\l_{\bar{1}\bar{2}}\l_{23}\l_{2\bar{3}}\l_{3\bar{3}}e^{2i(\cZ_1+\bar{\cZ}_1+\bar{\cZ}_2)} \\
& \quad +\l_1\bar{\l}_1\l_3\l_{1\bar{1}}\l_{13}\l_{\bar{1}3}\l_{2\bar{2}}\l_{2\bar{3}}\l_{\bar{2}\bar{3}}e^{2i(\cZ_1+\bar{\cZ}_1+\cZ_3)} +\l_1\bar{\l}_1\bar{\l}_3\l_{1\bar{1}}\l_{1\bar{3}}\l_{\bar{1}\bar{3}}\l_{2\bar{2}}\l_{23}\l_{\bar{2}3}e^{2i(\cZ_1+\bar{\cZ}_1+\bar{\cZ}_3)} \\
& \quad +\l_1\l_2\bar{\l}_2\l_{12}\l_{1\bar{2}}\l_{2\bar{2}}\l_{\bar{1}3}\l_{\bar{1}\bar{3}}\l_{3\bar{3}}e^{2i(\cZ_1+\cZ_2+\bar{\cZ}_2)} +\l_1\l_2\l_3\l_{12}\l_{13}\l_{23}\l_{\bar{1}\bar{2}}\l_{\bar{1}\bar{3}}\l_{\bar{2}\bar{3}}e^{2i(\cZ_1+\cZ_2+\cZ_3)} \\
& \quad +\l_1\l_2\bar{\l}_3\l_{12}\l_{1\bar{3}}\l_{2\bar{3}}\l_{\bar{1}\bar{2}}\l_{\bar{1}3}\l_{\bar{2}3}e^{2i(\cZ_1+\cZ_2+\bar{\cZ}_3)} +\l_1\bar{\l}_2\l_3\l_{1\bar{2}}\l_{13}\l_{\bar{2}3}\l_{\bar{1}2}\l_{\bar{1}\bar{3}}\l_{2\bar{3}}e^{2i(\cZ_1+\bar{\cZ}_2+\cZ_3)} \\
& \quad +\l_1\bar{\l}_2\bar{\l}_3\l_{1\bar{2}}\l_{1\bar{3}}\l_{\bar{2}\bar{3}}\l_{\bar{1}2}\l_{\bar{1}3}\l_{23}e^{2i(\cZ_1+\bar{\cZ}_2+\bar{\cZ}_3)}
+\l_1\l_3\bar{\l}_3\l_{13}\l_{1\bar{3}}\l_{3\bar{3}}\l_{\bar{1}2}\l_{\bar{1}\bar{2}}\l_{2\bar{2}}e^{2i(\cZ_1+\cZ_3+\bar{\cZ}_3)} \\
& \quad +\bar{\l}_1\l_2\bar{\l}_2\l_{\bar{1}2}\l_{\bar{1}\bar{2}}\l_{2\bar{2}}\l_{13}\l_{1\bar{3}}\l_{3\bar{3}}e^{2i(\bar{\cZ}_1+\cZ_2+\bar{\cZ}_2)} +\bar{\l}_1\l_2\l_3\l_{\bar{1}2}\l_{\bar{1}3}\l_{23}\l_{1\bar{2}}\l_{1\bar{3}}\l_{\bar{2}\bar{3}}e^{2i(\bar{\cZ}_1+\cZ_2+\cZ_3)} \\
& \quad +\bar{\l}_1\l_2\bar{\l}_3\l_{\bar{1}2}\l_{\bar{1}\bar{3}}\l_{2\bar{3}}\l_{1\bar{2}}\l_{13}\l_{\bar{2}3}e^{2i(\bar{\cZ}_1+\cZ_2+\bar{\cZ}_3)} +\bar{\l}_1\bar{\l}_2\l_3\l_{\bar{1}\bar{2}}\l_{\bar{1}3}\l_{\bar{2}3}\l_{12}\l_{1\bar{3}}\l_{2\bar{3}}e^{2i(\bar{\cZ}_1+\bar{\cZ}_2+\cZ_3)} \\
& \quad +\bar{\l}_1\bar{\l}_2\bar{\l}_3\l_{\bar{1}\bar{2}}\l_{\bar{1}\bar{3}}\l_{\bar{2}\bar{3}}\l_{12}\l_{13}\l_{23}e^{2i(\bar{\cZ}_1+\bar{\cZ}_2+\bar{\cZ}_3)} +\bar{\l}_1\l_3\bar{\l}_3\l_{\bar{1}3}\l_{\bar{1}\bar{3}}\l_{3\bar{3}}\l_{12}\l_{1\bar{2}}\l_{2\bar{2}}e^{2i(\bar{\cZ}_1+\cZ_3+\bar{\cZ}_3)} \\
& \quad +\l_2\bar{\l}_2\l_3\l_{2\bar{2}}\l_{23}\l_{\bar{2}3}\l_{1\bar{1}}\l_{1\bar{3}}\l_{\bar{1}\bar{3}}e^{2i(\cZ_2+\bar{\cZ}_2+\cZ_3)} +\l_2\bar{\l}_2\bar{\l}_3\l_{2\bar{2}}\l_{2\bar{3}}\l_{\bar{2}\bar{3}}\l_{1\bar{1}}\l_{13}\l_{\bar{1}3}e^{2i(\cZ_2+\bar{\cZ}_2+\bar{\cZ}_3)} \\
& \quad +\l_2\l_3\bar{\l}_3\l_{23}\l_{2\bar{3}}\l_{3\bar{3}}\l_{1\bar{1}}\l_{1\bar{2}}\l_{\bar{1}\bar{2}}e^{2i(\cZ_2+\cZ_3+\bar{\cZ}_3)} +\bar{\l}_2\l_3\bar{\l}_3\l_{\bar{2}3}\l_{\bar{2}\bar{3}}\l_{3\bar{3}}\l_{1\bar{1}}\l_{12}\l_{\bar{1}2}e^{2i(\bar{\cZ}_2+\cZ_3+\bar{\cZ}_3)},
\end{split}
\end{equation*}

\begin{footnotesize}
\begin{equation}
\begin{split}
N_2&=\l_3\bar{\l}_3\l_{1\bar{1}}\l_{12}\l_{1\bar{2}}\l_{\bar{1}2}\l_{\bar{1}\bar{2}}\l_{2\bar{2}}\l_{3\bar{3}}e^{2i(\cZ_1+\bar{\cZ}_1+\cZ_2+\bar{\cZ}_2)} +\bar{\l}_2\bar{\l}_3\l_{1\bar{1}}\l_{12}\l_{13}\l_{\bar{1}2}\l_{\bar{1}3}\l_{23}\l_{\bar{2}\bar{3}}e^{2i(\cZ_1+\bar{\cZ}_1+\cZ_2+\cZ_3)} \\
& \quad +\bar{\l}_2\l_3\l_{1\bar{1}}\l_{12}\l_{1\bar{3}}\l_{\bar{1}2}\l_{\bar{1}\bar{3}}\l_{2\bar{3}}\l_{\bar{2}3}e^{2i(\cZ_1+\bar{\cZ}_1+\cZ_2+\bar{\cZ}_3)} +
\l_2\bar{\l}_3\l_{1\bar{1}}\l_{1\bar{2}}\l_{13}\l_{\bar{1}\bar{2}}\l_{\bar{1}3}\l_{\bar{2}3}\l_{2\bar{3}}e^{2i(\cZ_1+\bar{\cZ}_1+\bar{\cZ}_2+\cZ_3)} \\
& \quad +\l_2\l_3\l_{1\bar{1}}\l_{1\bar{2}}\l_{1\bar{3}}\l_{\bar{1}\bar{2}}\l_{\bar{1}\bar{3}}\l_{\bar{2}\bar{3}}\l_{23}e^{2i(\cZ_1+\bar{\cZ}_1+\bar{\cZ}_2+\bar{\cZ}_3)} +
\l_2\bar{\l}_2\l_{1\bar{1}}\l_{13}\l_{1\bar{3}}\l_{\bar{1}3}\l_{\bar{1}\bar{3}}\l_{3\bar{3}}\l_{2\bar{2}}e^{2i(\cZ_1+\bar{\cZ}_1+\cZ_3+\bar{\cZ}_3)} \\
& \quad +\bar{\l}_1\bar{\l}_3\l_{12}\l_{1\bar{2}}\l_{13}\l_{2\bar{2}}\l_{23}\l_{\bar{2}3}\l_{\bar{1}\bar{3}}e^{2i(\cZ_1+\cZ_2+\bar{\cZ}_2+\cZ_3)} +\bar{\l}_1\l_3\l_{12}\l_{1\bar{2}}\l_{1\bar{3}}\l_{2\bar{2}}\l_{2\bar{3}}\l_{\bar{2}\bar{3}}\l_{\bar{1}3}e^{2i(\cZ_1+\cZ_2+\bar{\cZ}_2+\bar{\cZ}_3)} \\
& \quad +\bar{\l}_1\bar{\l}_2\l_{12}\l_{13}\l_{1\bar{3}}\l_{23}\l_{2\bar{3}}\l_{3\bar{3}}\l_{\bar{1}\bar{2}}e^{2i(\cZ_1+\cZ_2+\cZ_3+\bar{\cZ}_3)} +\bar{\l}_1\l_2\l_{1\bar{2}}\l_{13}\l_{1\bar{3}}\l_{\bar{2}3}\l_{\bar{2}\bar{3}}\l_{3\bar{3}}\l_{\bar{1}2}e^{2i(\cZ_1+\bar{\cZ}_2+\cZ_3+\bar{\cZ}_3)} \\
& \quad +\l_1\bar{\l}_3\l_{\bar{1}2}\l_{\bar{1}\bar{2}}\l_{\bar{1}3}\l_{2\bar{2}}\l_{23}\l_{\bar{2}3}\l_{1\bar{3}}e^{2i(\bar{\cZ}_1+\cZ_2+\bar{\cZ}_2+\cZ_3)} +\l_1\l_3\l_{\bar{1}2}\l_{\bar{1}\bar{2}}\l_{\bar{1}\bar{3}}\l_{2\bar{2}}\l_{2\bar{3}}\l_{\bar{2}\bar{3}}\l_{13}e^{2i(\bar{\cZ}_1+\cZ_2+\bar{\cZ}_2+\bar{\cZ}_3)} \\
& \quad +\l_1\bar{\l}_2\l_{\bar{1}2}\l_{\bar{1}3}\l_{\bar{1}\bar{3}}\l_{23}\l_{2\bar{3}}\l_{3\bar{3}}\l_{1\bar{2}}e^{2i(\bar{\cZ}_1+\cZ_2+\cZ_3+\bar{\cZ}_3)} +\l_1\l_2\l_{\bar{1}\bar{2}}\l_{\bar{1}3}\l_{\bar{1}\bar{3}}\l_{\bar{2}3}\l_{\bar{2}\bar{3}}\l_{3\bar{3}}\l_{12}e^{2i(\bar{\cZ}_1+\bar{\cZ}_2+\cZ_3+\bar{\cZ}_3)} \\
& \quad +\l_1\bar{\l}_1\l_{2\bar{2}}\l_{23}\l_{2\bar{3}}\l_{\bar{2}3}\l_{\bar{2}\bar{3}}\l_{3\bar{3}}\l_{1\bar{1}}e^{2i(\cZ_2+\bar{\cZ}_2+\cZ_3+\bar{\cZ}_3)},
\end{split}
\end{equation}
\end{footnotesize}
where $\l_{ij}\equiv\l_i-\l_j$, and $\mathcal{Z}_i=z/(\l_i-1)+\bar{z}/(\l_i+1)$.

\end{document}